# A combined model for the pseudorapidity distributions in *p-p* collisions at center-of-mass energies from 23.6 to 7000 GeV*

JIANG Zhi-Jin(姜志进)[†]   HUANG Yan(黄艳)   WANG Jie(王杰)

*College of Science, University of Shanghai for Science and Technology, Shanghai 200093, China*

In *p-p* collisions, the produced charge particles consist of two leading particles and those frozen out from the hot and dense matter created in collisions. The two leading particles are respectively in the projectile and target fragmentation region, which, in this paper, are conventionally supposed to have Gaussian rapidity distributions. The hot and dense matter is assumed to expand according to the unified hydrodynamics, a hydro model which unifies the features of Landau and Hwa-Bjorken model, and freeze out into charged particles from a space-like hypersurface with a fixed proper time of $\tau_{\text{FO}}$. The rapidity distribution of this part of charged particles can be derived out analytically. The combined contribution from both leading particles and unified hydrodynamics is then compared against the experimental data performed in a wide now available center-of-mass energy region from 23.6 to 7000 GeV. The model predictions are in well consistent with experimental measurements.

## 1 Introduction

The applications of relativistic hydrodynamics to depict the spatiotemporal evolution of matter created in *p-p* collisions may be traced back to the seminal work of Landau in 1953 [1]. However，due to the tiny collision system, low multiplicity and large fluctuation, the plausibility of this approach has met endless debates. At the same time, such strategy has achieved a great success in nucleus-nucleus collisions, where numerous theoretical and experimental researches have shown that the matter generated in such collisions behaves like a nearly ideal fluid [2-19].

The debating situation has changed since a series of findings of collective motion presented in *p-p* collisions in recent years [20-35], which is suitable for hydro description instead of others, such as Lund model that is ever a widely used theory in *p-p* collisions. Refs. [20-24] corroborated the formation of QGP (Quark Gluon Plasma) in *p-p* collisions, which is the basis for the practicability of hydrodynamics in such collisions. The measurements from CMS Collaboration at CERN-LHC have shown that [25], just like in nucleus-nucleus collisions, the ridge structures, the signal of collective motion, also appear in *p-p* collisions. Refs. [26, 27] further substantiated that these observed ridge structures can be well understood in the framework of hydrodynamics. The STAR Collaboration at BNL-RHIC has measured the HBT (Hanbury-Brown-Twiss) radii for *p-p* collisions as a function of multiplicity [28]. The measured results are well favored by the predictions of hydro models made in Res. [29-31]. In Refs. [32, 33], Sarkisyan *et al*. utilized the combined model of Landau hydrodynamics + combinations of the constituent quarks in participants to deal with the multihadron productions in *p-p* ($\bar{p}$) collisions. By employing the hydro solution known as Gubser flow [19], Ref. [34] demonstrated the collective radial flow of high multiplicity *p-p* collisions. In our previous work [35], we used the evolution-dominated hydrodynamics together with the effects of leading particles to describe the pseudorapidity distributions of charged particles produced in *p-p* collisions.

---

*partly supported by the Hujiang Foundation of China with Grant No. B14004; the Cultivating Subject of National Project with Grant No. 15HJPY-MS04.

[†]E-mail: jzj265@163.com

Pseudorapidity distribution is one of the most basic global variables. Compared with others, this variable has two main advantages. One is that it can be directly measured in experiments. This opens a way for understanding the properties of matter produced in collisions and the mechanism of particle production. The other is that it is only related to 1+1 flow of fluid. This simplifies the handling of problems greatly. In order to get pseudorapidity distributions, what we need is only to solve 1+1 dimensional equations. There is no necessary to deal with the complicated 3+1 expansions. In fact, owing to the tremendous complexity of hydro equations, from the times of Landau to now, what we can solve analytically is only limited to 1+1 flow for the ideal fluid with a simple equation of state. The 3+1 hydrodynamics is less developed, and no general exact solutions are known so far.

In the present paper, by taking into account the effect of leading particles, we shall discuss the pseudorapidity distributions in *p-p* collisions in the scope of unified hydrodynamics [9, 10], a hydro model which integrates the characteristics of Landau and Hwa-Bjorken model and can be solved analytically. Known from the investigations given in Ref. [10], this combined model is in good accordance with the experimental measurements carried out in heavy ion collisions from Cu-Cu to Pb-Pb at energies from RHIC to LHC scales. To clarify if and at what energy region this model is amenable to *p-p* collisions is the subject of this paper.

The organization of paper is as follows. The main points of combined model, including the exact solution of unified hydrodynamics and the rapidity distribution of charged particles frozen out from hot and dense matter, are listed in section 2. In section 3, a comparison is made between the model estimates and experimental data. A conclusion in section 4 closes the article.

## 2 A brief description of the combined model

## 1) The exact solution of unified hydrodynamics

The 1+1 flow of fluid follows the equation

$$\frac{\partial T^{\mu\nu}}{\partial x^{\nu}} = 0, \quad \mu, \nu = 0, 1, \tag{1}$$

where

$$T^{\mu\nu} = (\varepsilon + p)u^{\mu}u^{\nu} - pg^{\mu\nu} \tag{2}$$

is the energy-momentum tensor, $g^{\mu\nu} = \text{diag}(1,-1)$, $u^{\mu}$, $\varepsilon$ and $p$ are respectively the metric tensor, velocity, energy density and pressure of fluid. For a constant speed of sound, $\varepsilon$ and $p$ are related by the equation of state

$$\varepsilon = gp, \tag{3}$$

where $1/\sqrt{g} = c_s$ is the speed of sound, which in this paper takes the value of $c_s = 0.35$ [36]. Using Eqs. (2) and (3), and noticing the light-cone components of velocity

$$u_{\pm} = u^0 \pm u^1 = e^{\pm y},$$

where $y$ is the ordinary rapidity of fluid, Eq. (1) reads

$$\begin{aligned}
g\partial_+ \ln p &= -\frac{(g+1)^2}{2}\partial_+ y - \frac{g^2-1}{2}e^{-2y}\partial_- y, \\
g\partial_- \ln p &= \frac{(g+1)^2}{2}\partial_- y + \frac{g^2-1}{2}e^{2y}\partial_+ y,
\end{aligned} \tag{4}$$

where $\partial_+$ and $\partial_-$ are the compact notation of partial derivatives with respect to light-cone coordinates $z_{\pm} = t \pm z = x^0 \pm x^1 = \tau e^{\pm \eta_s}$, $t$ and $z$ are respectively the time and longitudinal coordinate, $\tau = \sqrt{z_+ z_-}$ is the proper time, and $\eta_s = 1/2 \ln(z_+/z_-)$ is the space-time rapidity of fluid.

The key ingredient of unified hydrodynamics is that it generalizes the relation between $y$ and $\eta_s$ by

$$2y = \ln u_+ - \ln u_- = \ln F_+(z_+) - \ln F_-(z_-), \tag{5}$$

where $F_{\pm}(z_{\pm})$ are a priori arbitrary function satisfying condition

$$F_{\pm} F''_{\pm} = \frac{A^2}{2}, \tag{6}$$

where $A$ is a constant. In case of $F_{\pm}(z_{\pm}) = z_{\pm}$, Eq. (5) reduces to $y = \eta_s$, returning to the boost-invariant picture of Hwa-Bjorken. Otherwise, Eq. (5) describes the non-boost-invariant geometry of Landau. Accordingly, Eq. (5) unifies the Hwa-Bjorken and Landau hydrodynamics together. It paves a way between these two models.

Substituting Eq. (5) into Eq. (4), we have

$$g\partial_+ \ln p = -\frac{(g+1)^2}{4}\frac{f'_+}{f_+} + \frac{g^2-1}{4}\frac{f'_-}{f_+},$$

$$g\partial_- \ln p = -\frac{(g+1)^2}{4}\frac{f'_-}{f_-} + \frac{g^2-1}{4}\frac{f'_+}{f_-},$$
(7)

where $f_\pm = F_\pm/H$, and $H$ is an arbitrary constant. The solution of above equation is

$$s(z_+,z_-) = s_0 \left(\frac{p}{p_0}\right)^{\frac{g}{g+1}} = s_0 \exp\left[-\frac{g+1}{4}(l_+^2+l_-^2) + \frac{g-1}{2}l_+l_-\right],$$
(8)

where $s$ is the entropy density of fluid, and

$$l_\pm(z_\pm) = \sqrt{\ln f_\pm},\quad y(z_+,z_-) = \frac{1}{2}(l_+^2 - l_-^2),\quad z_\pm = 2h\int_0^{l_\pm} e^{u^2} du.$$
(9)

where $h = H/A$.

**2) The rapidity distributions of charged particles frozen out from fluid**

By using Eq. (8), we can obtain the rapidity distributions of the charged particles frozen out from fluid or hot and dense matter created in collisions. To this end, we first evaluate the entropy distributions of the fluid on a space-like hypersurface with a fixed proper time of $\tau_{FO}$, from which the fluid will freeze out into the charged particles. Such distributions take the form as

$$\frac{dS}{dy} = su^\mu \left.\frac{d\lambda_\mu}{dy}\right|_{\tau_{FO}} = su^\mu n_\mu \left.\frac{d\lambda}{dy}\right|_{\tau_{FO}},$$
(10)

where $n^\mu$ is the 4-dimensional unit vector normal to the hypersurface

$$n^\mu n_\mu = n_+ n_- = 1.$$

$d\lambda^\mu = d\lambda n^\mu$, and $d\lambda$ is the space-like infinitesimal length element along hypersurface

$$d\lambda = \sqrt{d\lambda^\mu d\lambda_\mu} = \sqrt{-dz^+ dz^-}.$$
(11)

For a hypersurface with fixed proper time of $\tau_{FO}$, it can be taken as

$$\phi(z_+,z_-) = \tau_{FO}^2 = z_+ z_- = C.$$
(12)

where $C$ is a constant. Then the term on the right-hand side of Eq. (10)

$$u^\mu n_\mu d\lambda = \frac{1}{2}(u_+ z_- + u_- z_+)\frac{dz_-}{z_-} = \frac{1}{2}(u_+ z_- + u_- z_+)\frac{dz_+}{z_+}.$$
(13)

Furthermore, known from Eq. (9)

$$\mathrm{d}z_\pm = \frac{h}{l_\pm}\exp(l_\pm^2)\mathrm{d}l_\pm^2. \tag{14}$$

While, $\mathrm{d}l_\pm^2$ can be in turn written as

$$\mathrm{d}l_\pm^2 = \pm\frac{2z_\pm \exp(l_\mp^2)/l_\mp}{z_-\exp(l_+^2)/l_+ + z_+\exp(l_-^2)/l_-}\mathrm{d}y .$$

Thus, Eq. (14) becomes

$$\mathrm{d}z_\pm = \pm\frac{2hz_\pm}{z_-l_-\exp(-l_-^2) + z_+l_+\exp(-l_+^2)}\mathrm{d}y .$$

Substituting it into Eq. (13) and then into Eq. (10), we have

$$\frac{\mathrm{d}S}{\mathrm{d}y} = se^{\frac{1}{2}(l_+^2 + l_-^2)}\frac{z_-e^y + z_+e^{-y}}{z_-l_-e^y + z_+l_+e^{-y}}. \tag{15}$$

Inserting the entropy density of Eq. (8) into above equation and noticing the proportional relation between entropy and the number of charged particles, we have

$$\frac{\mathrm{d}N_{\mathrm{Fluid}}(\sqrt{s},y)}{\mathrm{d}y} = C(\sqrt{s})e^{-(g-1)(l_+ - l_-)^2/4}\frac{z_-e^y + z_+e^{-y}}{z_-l_-e^y + z_+l_+e^{-y}}, \tag{16}$$

where $C(\sqrt{s})$, independent of rapidity $y$, is an overall normalization constant. $\sqrt{s}$ is the center-of-mass energy.

**3) The rapidity distributions of leading particles**

In *p-p* collisions, there are two leading particles. One is in the projectile fragmentation region; the other is in the target fragmentation region. They also make contribution to the measured charged particles. As pointed out in Ref. [35], owing to the fact that, for a given incident energy, the leading particles in each time of *p-p* collisions have approximately the same amount of energy or rapidity, then，known from central limit theorem [37], the leading particles should possess Gaussian rapidity distribution, that is

$$\frac{\mathrm{d}N_{\mathrm{Lead}}(\sqrt{s},y)}{\mathrm{d}y} = \frac{1}{\sqrt{2\pi}\sigma}\exp\left\{-\frac{[|y| - y_0(\sqrt{s})]^2}{2\sigma^2}\right\}, \tag{17}$$

where $y_0$ and $\sigma$ are respectively the central position and width of distribution. It is obvious that $y_0$ should increase with incident energy. However, the width parameter $\sigma$, being relevant to the relative differences of rapidity among different leading particles, should not , at least not be apparently dependent on incident energy. Both the specific value of $y_0$ and $\sigma$ should now be determined by tuning the theoretical results to the experimental data.

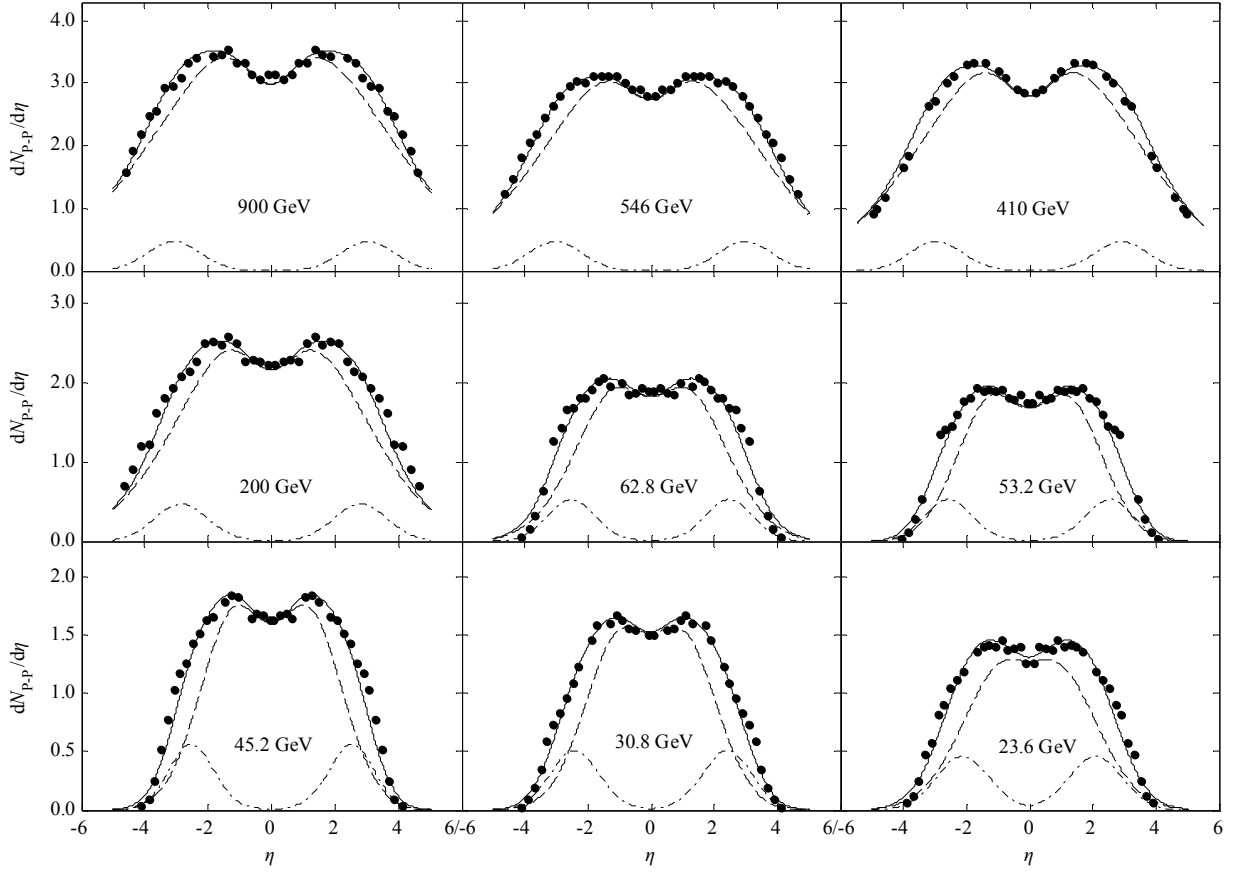

Fig. 1. The pseudorapidity distributions of charged particles produced in p-p collisions at $\sqrt{s}$ = 23.6 to 900 GeV, respectively. The solid dots are the experimental measurements [39-41]. The dashed curves are the results got from the unified hydrodynamics of Eq. (16). The dashed-dotted curves are the results obtained from the leading particles of Eq. (17). The solid curves are the results achieved from Eq. (20), that is, the sums of dashed and dashed-dotted curves.

## 3 Comparison with experimental measurements

From rapidity distributions, we can get pseudorapidity distributions by relation [38]

$$\frac{dN(\sqrt{s},\eta)}{d\eta} = \sqrt{1 - \frac{m^2}{m_T^2 \cosh^2 y}} \frac{dN(\sqrt{s},y)}{dy}, \quad (18)$$

where $m_T = \sqrt{m^2 + p_T^2}$ is the transverse mass, and $p_T$ is the transverse momentum. The first factor on the right-hand side of above equation is actually the Jacobian determinant. This transformation is closed by the following relation

$$y = \frac{1}{2}\ln\left[\frac{\sqrt{p_T^2 \cosh^2 \eta + m^2} + p_T \sinh \eta}{\sqrt{p_T^2 \cosh^2 \eta + m^2} - p_T \sinh \eta}\right]. \quad (19)$$

Taking into account the contributions from both the freeze-out of fluid and leading particles, the rapidity

distributions in Eq. 18 can be written as

$$\frac{dN(\sqrt{s},y)}{dy}=\frac{dN_{\text{Fluid}}(\sqrt{s},y)}{dy}+\frac{dN_{\text{Lead}}(\sqrt{s},y)}{dy}. \quad (20)$$

Inserting above equation or the sum of Eqs. (16) and (17) into (18), we can get the pseudorapidity distributions of the charged particles. Figure 1 shows such distributions in p-p collisions at energies from $\sqrt{s}$ =23.6 to 900 GeV. The solid dots in the figure are the experimental measurements [39-41]. The dashed curves are the results got from the unified hydrodynamics of Eq. (16). The dashed-dotted curves are the results obtained from the leading particles of Eq. (17). The solid curves are the results achieved from Eq. (20), that is, the sums of dashed and dashed-dotted curves. It can be seen that the theoretical results are well consistent with experimental measurements.

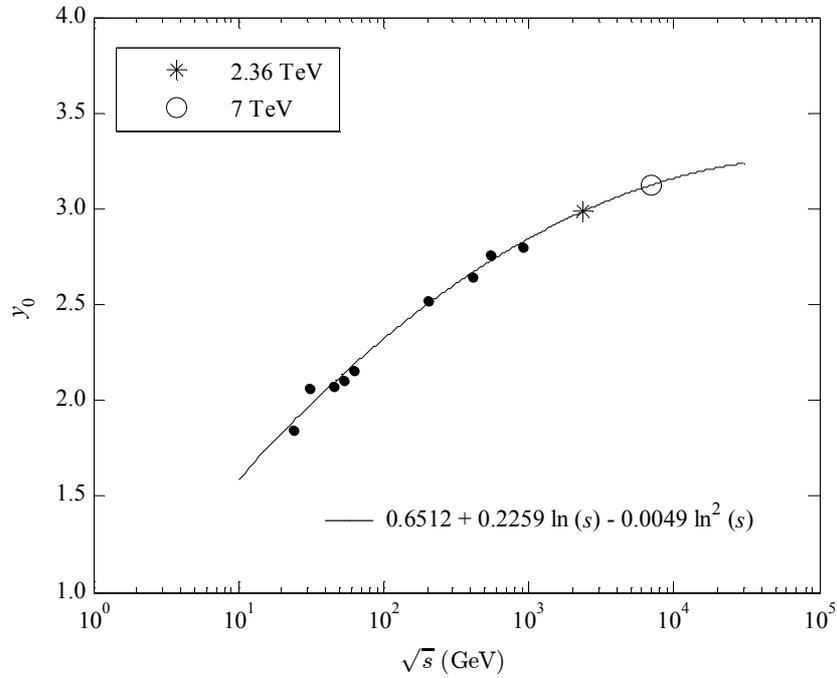

Fig. 2. The variation of $y_0$ against $\sqrt{s}$. The solid dots represent the fitted values listed in the text. The star and circle are the predictions for p-p collisions at CERN-LHC energies of $\sqrt{s}$ = 2.36 and 7 TeV, respectively. The solid curve is the result acquired from Eq. (22).

In calculations, the mass $m$ in Eqs. (18) and (19) takes the values of 0.23-0.32 GeV for energies from low to high, which are approximately the mean mass of pions, kaons and protons with the proportions of about 83%, 12% and 5% [42], respectively. The transverse momentum $p_T$ in Eqs. (18) and (19) takes the value according to the following formula extracted from experimental measurements [43]

$$p_T = 0.413 - 0.0171\ln(s) + 0.00143\ln^2(s), \quad (21)$$

where $p_T$ and $\sqrt{s}$ are respectively in the unit of GeV/c and GeV. The width parameter $\sigma$ in Eq. (17) takes a constant of $\sigma = 0.9$ in the whole energy region. As the analyses given above, it is independent of energy. The center parameter $y_0$ in Eq. (17) takes the values of 1.85, 2.06, 2.07, 2.10, 2.15, 2.52, 2.65, 2.76, 2.78 for energies from 23.6 to 900 GeV. As mentioned above, it increases with energy. Its variation against energy is shown in Fig. 2. The solid curve in this figure meets relation

$$y_0 = 0.6512 + 0.2259\ln(s) - 0.0049\ln^2(s), \tag{22}$$

where $\sqrt{s}$ is in the unit of GeV. It can be seen that all the solid dots representing the above fitted values are well seated on the curve. The star and circle stand for the values of *p-p* collisions at CERN-LHC energies of $\sqrt{s} = 2.36$ and 7 TeV, respectively, which allow us to predict the pseudorapidity distributions of charged particles in these two cases, and the results are shown in Fig. 3. The solid dots are the experimental data [43], and the meanings of different curves are the same as those in Fig. 1. It can be seen that the theoretical results favor well with the available measurements in the mid-pseudorapidity regions.

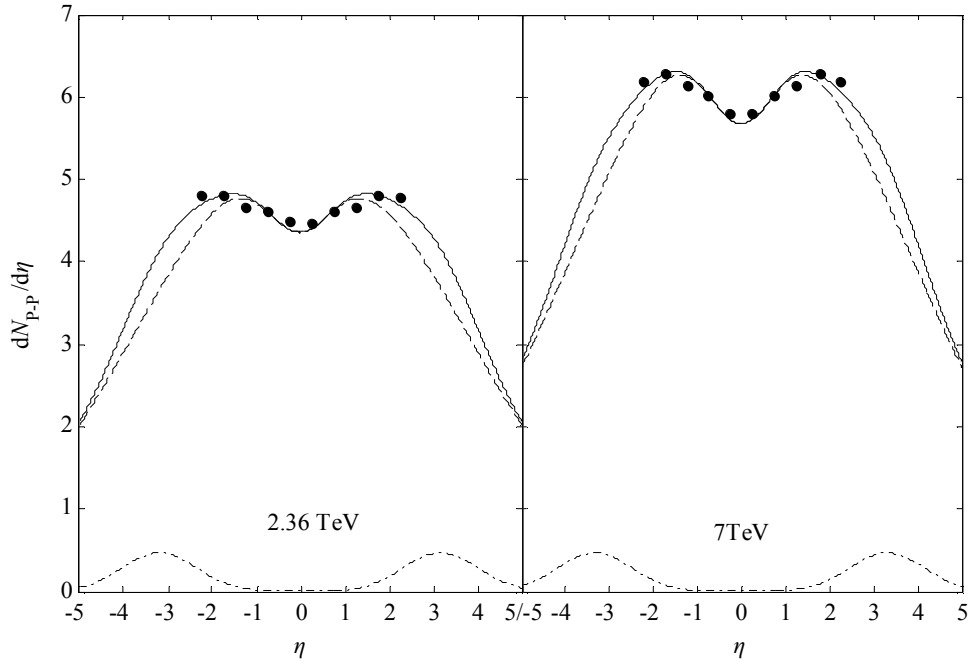

Fig. 3. The predicted pseudorapidity distributions of charged particles produced in *p-p* collisions at CERN-LHC energies of $\sqrt{s} = 2.36$ and 7 TeV, respectively. The solid dots are the experimental measurements [43]. The dashed curves are the results from the unified hydrodynamics of Eq. (16). The dashed-dotted curves are the results from the leading particles of Eq. (17). The solid curves are the results achieved from Eq. (20), that is, the sums of dashed and dashed-dotted curves.

## 4 Conclusions

In *p-p* collisions, there are only two leading particles which are separately in projectile and target fragmentation region. As before, these leading particles are assumed having Gaussian rapidity distributions in view of the fact that the leading particles formed in different collisions share approximately the same energy.

The matter created in collisions is assumed evolving according to unified hydrodynamics which, by generalizing the relation between ordinary rapidity $y$ and space-time one $\eta_s$, makes the Hwa-Bjorken and Landau two famous hydro models come together. In case of linear equation of state, this hydro model can be solved analytically and the solution can be used to formulate the rapidity distributions of charged particles frozen out from the fluid at the space-like hypersurface with a fixed proper time of $\tau_{FO}$. Known from comparing with experimental data, the total contributions from both unified hydrodynamics and leading particles match up well with the observations in the energy region from $\sqrt{s}$ =23.6 to 900 GeV. Using the predicted parameters, the combined model also describes well the experimental measurements, at least in the mid-pseudorapidity region, for the incident energies up to the CERN-LHC scale of 2.36 and 7 TeV.

In our previous work [35], by taking into account the effect of leading particles, we once used the evolution-dominated hydrodynamics to successfully describe part of experimental data adopted here. This hydro model differs from the one used in this paper in two ways: (1) Different initial conditions. The former assumes that the fluid is initially at rest. Its motion is totally dominated by the following evolution. The latter, as stated above, employs the initial condition of Eq. (5). It plays a connection between Hwa-Bjorken and Landau hydrodynamics. (2) Different freeze-out conditions. The former assumes that the freeze-out of fluid occurs at the space-like hypersurface with a fixed temperature of $T_{FO}$. The latter, as aforementioned, takes such hypersurface as one with a fixed proper time of $\tau_{FO}$. It is interesting to notice that, in both cases, the width parameter $\sigma$ in Eq. (17) takes the same value of $\sigma = 0.9$. The two sets of fitted center parameter $y_0$ in this equation also have no much difference.

The existing studies show that both theoretical models can give an equally proper description to the experimental observations. Then the question arises: which model is more reasonable? This judgment might remain to the experimental or further theoretical investigations. Only after the free parameters in models, such as the speed of sound $\sqrt{c_s}$ and the center parameter $y_0$, are definitely determined in experiment or theory can the answer become clear. In the absence of further experimental or theoretical evidences, every theoretical model might be potentially a good one if it is rational enough in theory and not contradict to the present experimental data.